\documentclass[12pt]{article}
\usepackage{amssymb}

\input{tcilatex}
\begin{document}

\author{Craig Liu and Deng-Feng Wang \\
%EndAName
Centre of Advanced Studies in Finance \\
University of Waterloo\\
Waterloo, ON N2L 3E5 Canada}
\title{\textbf{Exchange options and spread options }\\
\textbf{\ with stochastic interest rates}}
\date{Mar. 1, 1999}
\maketitle

\begin{abstract}
In this work, we consider the issue of pricing exchange options and spread
options with stochastic interest rates. We provide the closed form solution
for the exchange option price when interest rate is stochastic. Our result
holds when interest rate is modeled with a stochastic term structure of
general form, which includes Vasicek model, CIR term structure, and other
well-known term structure models as special cases. In particular, we have
discussed the possibility of using our closed form solution as a control
variate in pricing spread options with stochastic interest rate.
\end{abstract}

\newpage

\section{Introduction}

Spread options have become increasingly important. They give the holders the
right to call or put the spread value of two underlying assets against a
predetermined parameter $K$ as the strike price. In particular, the spread
options reduce to the so-called exchange options when the predetermined
strike price $K$ is set to zero. Spread options and exchange options can be
viewed as options to exchange one underlying asset for another with respect
to the strike price. They are used in many situations. One typical example
is that the option holder is interested in exchanging one commodity for
another commodity. For instance, in oil industries, the prices of crude oil
and refined oil differ from each other, and both prices are fluctuating
considerably in response to the weather, regional stabilities of world oil
production centers, and other human and natural parameters. Oil companies
may deal with the situations of price fluctuations using the spread options
or exchange options. Spread and exchange options have been of considerable
interests to both practitioners and theoretical researchers\cite{marg,garman,tan}.

For the exchange option, a closed form solution for its price is available.
The valuation of exchange option was first 
studied by Margrabe\cite{marg}, based
on the option pricing theory of Black and Scholes\cite{black}, and Merton\cite{merton1,merton2}.
The derivation of Margrabe is a PDE approach\cite{marg}. 
However, in Margrabe's
derivation, the risk-free rate $r$ is assumed to be a constant, which is far
from reality. It is of great interest to both practitioners and theoretical
researchers to investigate whether closed form solution exists when the
interest rate is modeled with stochastic term structure.

This paper investigates how stochastic interest rate will affect the
exchange option pricing. The closed form solution for the exchange option's
price is given when we assume a very general stochastic process for the
interest rate, which includes Vasicek model\cite{vasicek}, 
CIR model\cite{cir}, affine term
structure models\cite{kan}
and other interest rate models as special cases. To our
knowledge, this is the first time to provide the closed form solution for
exchange option pricing while stochastic interest rate is taken into account.

We also argue that to price a European style spread option of general strike
price $K$, one may use our closed form solution as a control variate to
reduce variance of simulation when doing Monte Carlo pricing\cite{phelim}
for stochastic interest rate. The closed
form result presented here shall be of interest to both theoretician and
practitioners.

In our discussion below, we assume an exchange economy populated by
risk-averse agents with increasing preferences, and all economic activities
take place in the time interval $[0,T]$. All the possible outcomes of
this economy is denoted by a measurable space $(\Omega ,\mathcal{F})$ where $%
\Omega $ is the set of all possible states and $\mathcal{F}$ is a sigma
algebra of subsets of $\Omega .$ Information arrival in this economy is
described by a filtration $\{\mathcal{F}_{t};0\leq t\leq T\}$ with $\mathcal{%
F}_{T}=\mathcal{F},$ and the agents belief is modelled by a probability
measure $P$ defined on $(\Omega ,\mathcal{F}).$

In following sections, we will first rederive the closed form solution of
exchange option using the risk-neutral martingale approach. It is shown that
this provides identical result to the one given by Margrabe using partial
differential equation. Then, we discuss, within the framework of martingale
measure, how to price the exchange option when interest rate is stochastic.
It is shown that our result is valid for most general term structure, only
the correlation coefficients between the interest rate and the underlying
assets will affect the option price.

\section{Exchange Options}

Exchange options can be defined by using two underlying assets or
commodities which are closely related. This correlation between the two
assets or commodities results from demand substitution or the potential for
transformation. In general, an exchange option has the following payoff: 
\begin{equation}
\max \{\lambda (S_{1}(T)-S_{2}(T)),0\}=\left[ \lambda \left(
S_{1}(T)-S_{2}(T)\right) \right] ^{+}  \label{a1}
\end{equation}
where $\lambda =1$ for a call and $\lambda =-1$ for a put, and $S_{1}(T)$
and $S_{2}(T)$ are the underlying asset prices at maturity $T$.

\subsection{Constant interest rate}

Let us first review the case of constant interest rate. The pricing closed
form was given by Margrabe via PDE method. In the following, we give a short
review within the framework of risk neutral
martingales\cite{cox, duffie,har1,har2}.
Part of the results will be used in the section of discussing stochastic interest rate case.

Suppose that in the physical probability space $(\Omega ,\mathcal{F},P)$,
the prices of two underlying assets for an exchange option follow the
geometric Brownian motions, that is, 
\begin{equation}
\frac{dS_{i}(t)}{S_{i}(t)}=\mu _{i}dt+\sigma _{i}dW_{i}(t),\qquad i=1,\text{ 
}2  \label{ee5}
\end{equation}
where $\mu _{1},$ $\mu _{1},$ $\sigma _{1},$ $\sigma _{2},$ and $\rho
_{12}\equiv Corr[dW_{1},dW_{2}]$ are all constants. The price $%
C(t,S_{1}(t),S_{2}(t))$ of a (European) call exchange option at time $t$ is
then given by 
\begin{equation}
C(t,S_{1}(t),S_{2}(t))=E_{t}^{Q}\left[ e^{-\int_{t}^{T}r(s)ds}\ \left[
S_{1}(T)-S_{2}(T)\right] ^{+}\right]  \label{ee6}
\end{equation}
where $Q$ is the corresponding equivalent martingale measure. We should note
here that these two processes in (\ref{ee5}) are defined in the physical
probability space $(\Omega ,\mathcal{F},P),$ however, the general valuation
formula in (\ref{ee6}) is derived in the risk-neutral probability space $%
(\Omega ,\mathcal{F},Q).$ The relationship between the risk-neutral
probability space and the physical probability space is the standard one,
which is described by the Girsanov transformation\cite{duffie}.

If the risk-free rate $r$ is constant, then the closed form formula for
pricing the exchange option was first derived by Margrabe (1978) using the
partial differential equation approach. However, we can also compute the
expectation value in the risk neutral space, and the exchange option price
will follow. It is shown that the approach gives the pricing formula
identical to the one derived by Margrabe. In this case, we assume that there
is no dividend paying during the option's life. Since the price processes of
the two underlying assets are governed by (\ref{ee5}), under the equivalent
martingale measure $Q$ with constant risk-free rate $r$, we then have 
\begin{equation}
\left[ \left. 
\begin{array}{c}
\log [S_{1}(T)] \\ 
\log [S_{2}(T)]
\end{array}
\right| \mathcal{F}_{t}\right] \stackrel{D}{\thicksim }\mathcal{N}\left(
\left[ 
\begin{array}{c}
A_{1} \\ 
A_{2}
\end{array}
\right] ,\left[ 
\begin{array}{cc}
\nu _{1}^{2} & \rho _{12}\nu _{1}\nu _{2} \\ 
\rho _{12}\nu _{1}\nu _{2} & \nu _{2}^{2}
\end{array}
\right] \right)  \label{qq3}
\end{equation}
where 
\begin{equation}
A_{i}=\log [S_{i}(t)]+(r-\frac{\sigma _{i}^{2}}{2})(T-t),\qquad \nu
_{i}^{2}=\sigma _{i}^{2}(T-t),\qquad i=1,\text{ }2.  \label{qq4}
\end{equation}
Note that we can write 
\[
S_{1}(T)-S_{2}(T)=e^{A_{1}+\sigma _{1}\sqrt{T-t}Z_{1}}\left\{
1-e^{A_{2}-A_{1}+\sigma _{2}\sqrt{T-t}Z_{2}-\sigma _{1}\sqrt{T-t}%
Z_{1}}\right\} 
\]
where 
\[
\left[ 
\begin{array}{c}
Z_{1} \\ 
Z_{2}
\end{array}
\right] \stackrel{D}{\thicksim }\mathcal{N}\left( \left[ 
\begin{array}{c}
0 \\ 
0
\end{array}
\right] ,\left[ 
\begin{array}{ll}
1 & \rho _{12} \\ 
\rho _{12} & 1
\end{array}
\right] \right) . 
\]
Hence, 
\[
S_{1}(T)\geq S_{2}(T)\Leftrightarrow Z_{3}\geq m 
\]
where 
\begin{eqnarray*}
Z_{3} &\equiv &\frac{\sigma _{1}Z_{1}-\sigma _{2}Z_{2}}{\sqrt{(\sigma
_{1}^{2}+\sigma _{2}^{2}-2\rho _{12}\sigma _{1}\sigma _{2})}}, \\
m &\equiv &\frac{A_{2}-A_{1}}{\sqrt{(\sigma _{1}^{2}+\sigma _{2}^{2}-2\rho
_{12}\sigma _{1}\sigma _{2})(T-t)}},
\end{eqnarray*}
and 
\[
\left[ 
\begin{array}{c}
Z_{1} \\ 
Z_{3}
\end{array}
\right] \stackrel{D}{\thicksim }\mathcal{N}\left( \left[ 
\begin{array}{c}
0 \\ 
0
\end{array}
\right] ,\left[ 
\begin{array}{ll}
1 & \eta \\ 
\eta & 1
\end{array}
\right] \right) 
\]
with 
\[
\eta \equiv \frac{\sigma _{1}-\sigma _{2}\rho _{12}}{\sqrt{\sigma
_{1}^{2}+\sigma _{2}^{2}-2\rho _{12}\sigma _{1}\sigma _{2}}}. 
\]
So, 
\begin{eqnarray}
&&E_{t}^{Q}\left[ \left[ S_{1}(T)-S_{2}(T)\right] ^{+}\right]  \nonumber \\
&=&\int_{-\infty }^{+\infty }\int_{m}^{+\infty }e^{A_{1}+\sigma _{1}\sqrt{T-t%
}x}p(x,y,\eta )dydx-\int_{-\infty }^{+\infty }\int_{m}^{+\infty
}e^{A_{2}+\sigma _{1}\sqrt{T-t}x+by}p(x,y,\eta )dydx  \nonumber \\
&\equiv &I_{1}-I_{2}  \label{ii}
\end{eqnarray}
where 
\begin{eqnarray*}
b &\equiv &\sqrt{(\sigma _{1}^{2}+\sigma _{2}^{2}-2\rho _{12}\sigma
_{1}\sigma _{2})(T-t)}, \\
p(x,y,\eta ) &\equiv &\frac{1}{2\pi \sqrt{1-\eta ^{2}}}e^{-\frac{1}{2(1-\eta
^{2})}(x^{2}-2\eta xy+y^{2})}.
\end{eqnarray*}
Note that 
\begin{eqnarray}
I_{1} &=&\int_{-\infty }^{+\infty }\left\{ \int_{m}^{+\infty
}e^{A_{1}+\sigma _{1}\sqrt{T-t}x}p(x,y,\eta )dy\right\} dx  \nonumber \\
&=&\int_{m}^{+\infty }\frac{1}{\sqrt{2\pi }}e^{A_{1}-\frac{1}{2(1-\eta ^{2})}%
\left( y^{2}-\left( \eta y+(1-\eta ^{2})\sigma _{1}\sqrt{T-t}\right)
^{2}\right) }dy  \nonumber \\
&=&e^{A_{1}+\frac{1}{2}(1-\eta ^{2})\sigma _{1}^{2}(T-t)+\eta ^{2}\sigma
_{1}^{2}(T-t)}\int_{m}^{+\infty }\frac{1}{\sqrt{2\pi }}e^{-\frac{1}{2}\left(
y-\eta \sigma _{1}\sqrt{T-t}\right) ^{2}}dy  \nonumber \\
&=&e^{A_{1}+\frac{1}{2}\sigma _{1}^{2}(T-t)}\Phi (\eta \sigma _{1}\sqrt{T-t}%
-m)  \label{i1}
\end{eqnarray}
and 
\begin{eqnarray}
I_{2} &=&\int_{-\infty }^{+\infty }\left\{ \int_{m}^{+\infty
}e^{A_{2}+\sigma _{1}\sqrt{T-t}x-by}p(x,y,\eta )dy\right\} dx  \nonumber \\
&=&\int_{m}^{+\infty }e^{A_{2}-by}\frac{1}{\sqrt{2\pi }}e^{-\frac{1}{%
2(1-\eta ^{2})}\left[ y^{2}-\left( \eta y+(1-\eta ^{2})\sigma _{1}\sqrt{T-t}%
\right) ^{2}\right] }dy  \nonumber \\
&=&e^{A_{2}+\frac{1}{2}\sigma _{1}^{2}(T-t)(1-\eta ^{2})+\frac{1}{2}(\eta
\sigma _{1}\sqrt{T-t}-b)^{2}}\int_{m}^{+\infty }\frac{1}{\sqrt{2\pi }}e^{-%
\frac{1}{2}\left( y-\eta \sigma _{1}\sqrt{T-t}+b\right) ^{2}}dy  \nonumber \\
&=&e^{A_{2}+\frac{1}{2}\sigma _{1}^{2}(T-t)-\eta b\sigma _{1}\sqrt{T-t}+%
\frac{b^{2}}{2}}\Phi (\eta \sigma _{1}\sqrt{T-t}-m-b)  \label{i2}
\end{eqnarray}
where $\Phi (\cdot )$ is the standard Gaussian distribution function.
Therefore, by (\ref{ii}), (\ref{i1}), and (\ref{i2}), we obtain 
\begin{eqnarray*}
&&E_{t}^{Q}\left[ e^{-r(T-t)}\left[ S_{1}(T)-S_{2}(T)\right] ^{+}\right] \\
&=&S_{1}(t)\Phi (\eta \sigma _{1}\sqrt{T-t}-m)-S_{2}(t)\Phi (\eta \sigma _{1}%
\sqrt{T-t}-m-b).
\end{eqnarray*}
To sum up, for constant interest rate $r$, the price of a call exchange
option at time $t$, is given by 
\begin{equation}
E_{t}^{Q}\left[ e^{-r(T-t)}\left[ S_{1}(T)-S_{2}(T)\right] ^{+}\right]
=S_{1}(t)\Phi (d_{1})-S_{2}(t)\Phi (d_{2})  \label{qq0}
\end{equation}
where 
\begin{eqnarray}
d_{1} &=&\frac{\log [S_{1}(t)/S_{2}(t)]}{\sqrt{(\sigma _{1}^{2}+\sigma
_{2}^{2}-2\rho _{12}\sigma _{1}\sigma _{2})(T-t)}}+\frac{1}{2}\sqrt{(\sigma
_{1}^{2}+\sigma _{2}^{2}-2\rho _{12}\sigma _{1}\sigma _{2})(T-t)},
\label{qq1} \\
d_{2} &=&d_{1}-\sqrt{(\sigma _{1}^{2}+\sigma _{2}^{2}-2\rho _{12}\sigma
_{1}\sigma _{2})(T-t)},  \label{qq2}
\end{eqnarray}
and $\Phi (\cdot )$ is the standard Gaussian distribution function. This is
consistent with the result of Margrabe derived with partial differential
equation approach\cite{marg}.

As shown above, the exchange option pricing can also be obtained within the
framework of the risk neutral measure, consistent with the result obtained
by Margrabe, which was derived with partial differential equation approach.
In this case of constant interest rate, we wish to note that the interest
rate $r$ does not enter the pricing formula explicitly. This special feature
motivates us to look into the issue of pricing exchange option when interest
rate is stochastic in a general form. The next subsection discusses this in
details within the framework of risk-neutral measures.

\subsection{Stochastic interest rate}

In this subsection, we discuss the issue of pricing exchange options when
interest rate is stochastic. It will be shown below that the option price
closed form can be found for most general one-factor stochastic interest
rate processes ( i.e. one Wiener process ). Our result applies to the cases
where one describes the interest rate such as Vasicek term structure, CIR
term structure. These are special cases of our consideration.

In the following, it is assumed that we are always working in the
risk-neutral probability space $(\Omega ,\mathcal{F},Q)$. Each process below
is referred to this risk-neutral probability measure $Q$. Now assume that
the short rate $r$ also follows a Markov diffusion process, that is, 
\[
dr(t)=\mu (r(t),t)dt+\sigma (r(t),t)dW_{0}(t). 
\]
Here, we do not specify a concrete interest rate model. All we need to
assume is that the interest rate is a Markov diffusion process. The interest
rate is correlated with the two underlying assets of the exchange option
being considered. Assume further that the correlation matrix of $%
[dW_{0}(t),dW_{1}(t),dW_{2}(t)]$ is 
\[
\left[ 
\begin{array}{lll}
1 & \rho _{01} & \rho _{02} \\ 
\rho _{01} & 1 & \rho _{12} \\ 
\rho _{02} & \rho _{12} & 1
\end{array}
\right] 
\]
where $\rho _{01},$ $\rho _{02},$ and $\rho _{12}$ are constants. Using
Cholesky decomposition, we can obtain the above correlation structure by
setting 
\begin{equation}
\left\{ 
\begin{array}{l}
dW_{0}(t)=dB_{0}(t) \\ 
dW_{1}(t)=\rho _{01}dB_{0}(t)+\sqrt{1-\rho _{01}^{2}}dB_{1}(t) \\ 
dW_{2}(t)=\rho _{02}dB_{0}(t)+\frac{\rho _{12}-\rho _{01}\rho _{02}}{\sqrt{%
1-\rho _{01}^{2}}}dB_{1}(t)+\sqrt{1-\rho _{02}^{2}-\frac{(\rho _{12}-\rho
_{01}\rho _{02})^{2}}{1-\rho _{01}^{2}}}dB_{2}(t)
\end{array}
\right.  \label{e2}
\end{equation}
where $B_{0}(t),$ $B_{1}(t),$ and $B_{2}(t)$ are three independent standard
Brownian motions. For the underlying assets, their prices will follow the
processes below: 
\begin{eqnarray*}
\log [S_{1}(T)] &=&\tilde{A}_{1}+\tilde{\sigma}_{1}\int_{t}^{T}dB_{1}(s) \\
\log [S_{2}(T)] &=&\tilde{A}_{2}+\tilde{\sigma}_{2}\left[ \tilde{\rho}%
_{12}\int_{t}^{T}dB_{1}(s)+\sqrt{1-\tilde{\rho}_{12}^{2}}%
\int_{t}^{T}dB_{2}(s)\right]
\end{eqnarray*}
where 
\begin{eqnarray}
\tilde{A}_{1} &=&\log [S_{1}(t)]+\int_{t}^{T}r(s)ds-\frac{1}{2}\sigma
_{1}^{2}(T-t)+\sigma _{1}\rho _{01}\int_{t}^{T}dB_{0}(s),  \nonumber \\
\tilde{A}_{2} &=&\log [S_{2}(t)]+\int_{t}^{T}r(s)ds-\frac{1}{2}\sigma
_{2}^{2}(T-t)+\sigma _{2}\rho _{02}\int_{t}^{T}dB_{0}(s),  \nonumber \\
\tilde{\sigma}_{1} &=&\sigma _{1}\sqrt{1-\rho _{01}^{2}},  \label{ss1} \\
\tilde{\sigma}_{2} &=&\sigma _{2}\sqrt{1-\rho _{02}^{2}},  \label{ss2} \\
\tilde{\rho}_{12} &=&\frac{\rho _{12}-\rho _{01}\rho _{02}}{\sqrt{1-\rho
_{01}^{2}}\sqrt{1-\rho _{02}^{2}}}.  \label{ss3}
\end{eqnarray}
And hence, by conditional expectation, we can price a call exchange option
as 
\begin{eqnarray}
&&E_{t}^{Q}\left[ e^{-\int_{t}^{T}r(s)ds}\ \left[ S_{1}(T)-S_{2}(T)\right]
^{+}\right]  \nonumber \\
&=&E_{t}^{Q}\left[ e^{-\int_{t}^{T}r(s)ds}E_{t}^{Q}\left( \left[
S_{1}(T)-S_{2}(T)\right] ^{+}\left| \{B_{0}(s):t\leq s\leq T\}\right.
\right) \right] .  \label{ex1}
\end{eqnarray}
Note that given a sample path of $\{B_{0}(s):t\leq s\leq T\}$, using the
results in (\ref{i1}) and (\ref{i2}), we obtain the following: 
\begin{eqnarray}
&&E_{t}^{Q}\left( \left[ S_{1}(T)-S_{2}(T)\right] ^{+}\left|
\{B_{0}(s):t\leq s\leq T\}\right. \right)  \label{ex2} \\
&=&e^{\tilde{A}_{1}+\frac{1}{2}\tilde{\sigma}_{1}^{2}(T-t)}\Phi (\tilde{\eta}%
\tilde{\sigma}_{1}\sqrt{T-t}-\tilde{m})-e^{\tilde{A}_{2}+\frac{1}{2}\tilde{%
\sigma}_{1}^{2}(T-t)-\tilde{\eta}\tilde{b}\tilde{\sigma}_{1}\sqrt{T-t}+\frac{%
\tilde{b}^{2}}{2}}\Phi (\tilde{\eta}\tilde{\sigma}_{1}\sqrt{T-t}-\tilde{m}-%
\tilde{b})  \nonumber
\end{eqnarray}
where 
\begin{eqnarray*}
\tilde{\eta} &\equiv &\frac{\tilde{\sigma}_{1}-\tilde{\sigma}_{2}\tilde{\rho}%
_{12}}{\sqrt{\tilde{\sigma}_{1}^{2}+\tilde{\sigma}_{2}^{2}-2\tilde{\rho}_{12}%
\tilde{\sigma}_{1}\tilde{\sigma}_{2}}}, \\
\tilde{m} &\equiv &\frac{\tilde{A}_{2}-\tilde{A}_{1}}{\sqrt{(\tilde{\sigma}%
_{1}^{2}+\tilde{\sigma}_{2}^{2}-2\tilde{\rho}_{12}\tilde{\sigma}_{1}\tilde{%
\sigma}_{2})(T-t)}}, \\
\tilde{b} &\equiv &\sqrt{(\tilde{\sigma}_{1}^{2}+\tilde{\sigma}_{2}^{2}-2%
\tilde{\rho}_{12}\tilde{\sigma}_{1}\tilde{\sigma}_{2})(T-t)}.
\end{eqnarray*}
And it is straightforward to check 
\begin{eqnarray}
\tilde{\eta}\tilde{\sigma}_{1}\sqrt{T-t}-\tilde{m} &=&\frac{\log
[S_{1}(t)/S_{2}(t)]+\frac{T-t}{2}(\sigma _{2}^{2}\rho _{02}^{2}-\sigma
_{1}^{2}\rho _{01}^{2})+(\sigma _{1}\rho _{01}-\sigma _{2}\rho
_{02})\int_{t}^{T}dB_{0}(s)}{\sqrt{(\tilde{\sigma}_{1}^{2}+\tilde{\sigma}%
_{2}^{2}-2\tilde{\rho}_{12}\tilde{\sigma}_{1}\tilde{\sigma}_{2})(T-t)}} 
\nonumber \\
&&+\frac{1}{2}\sqrt{(\tilde{\sigma}_{1}^{2}+\tilde{\sigma}_{2}^{2}-2\tilde{%
\rho}_{12}\tilde{\sigma}_{1}\tilde{\sigma}_{2})(T-t)}, \\
\tilde{\sigma}_{1}^{2}+\tilde{\sigma}_{2}^{2}-2\tilde{\rho}_{12}\tilde{\sigma%
}_{1}\tilde{\sigma}_{2} &=&(\sigma _{1}^{2}+\sigma _{2}^{2}-2\rho
_{12}\sigma _{1}\sigma _{2})-(\sigma _{1}\rho _{01}-\sigma _{2}\rho
_{02})^{2}.  \label{ex4} \\
e^{\tilde{A}_{1}+\frac{1}{2}\tilde{\sigma}_{1}^{2}(T-t)}
&=&S_{1}(t)e^{\sigma _{1}\rho _{01}x-\frac{1}{2}\sigma _{1}^{2}\rho
_{01}^{2}(T-t)},  \label{ex5} \\
e^{\tilde{A}_{2}+\frac{1}{2}\tilde{\sigma}_{1}^{2}(T-t)-\tilde{\eta}\tilde{b}%
\tilde{\sigma}_{1}\sqrt{T-t}+\frac{\tilde{b}^{2}}{2}} &=&S_{2}(t)e^{\sigma
_{2}\rho _{02}x-\frac{1}{2}\sigma _{2}^{2}\rho _{02}^{2}(T-t)}.  \label{ex6}
\end{eqnarray}

Since $\int_{t}^{T}dB_{0}(s)\left| \mathcal{F}_{t}\right. \stackrel{D}{%
\thicksim }\mathcal{N}\left( 0,T-t\right) ,$ by (\ref{ex1}), $\cdots ,$ (\ref
{ex6}), we then obtain exchange option price with stochastic interest rates:
Under the conditions $(C1)$ the prices of two underlying assets follow the
geometric Brownian motions (\ref{ee5}), and there is no dividend paying
during the option's life; $(C2)$ If the risk-free rate $r$ is stochastic$,$
then the price of a call exchange option at time $t$, defined by (\ref{ee6}%
), is given by 
\begin{eqnarray*}
&&E_{t}^{Q}\left[ e^{-\int_{t}^{T}r(s)ds}\ \left[ S_{1}(T)-S_{2}(T)\right]
^{+}\right]  \\
&=&S_{1}(t)\int_{-\infty }^{+\infty }\ \phi (x;\sigma _{1}\rho
_{01},T-t)\Phi (d_{1}(x))dx-S_{2}(t)\int_{-\infty }^{+\infty }\phi (x;\sigma
_{2}\rho _{02},T-t)\Phi (d_{2}(x))\ dx
\end{eqnarray*}
where 
\begin{eqnarray*}
d_{1}(x) &=&\frac{\log \left[ S_{1}(t)/S_{2}(t)\right] +\frac{T-t}{2}(\sigma
_{2}^{2}\rho _{02}^{2}-\sigma _{1}^{2}\rho _{01}^{2})+(\sigma _{1}\rho
_{01}-\sigma _{2}\rho _{02})x}{\sqrt{\left[ (\sigma _{1}^{2}+\sigma
_{2}^{2}-2\rho _{12}\sigma _{1}\sigma _{2})-(\sigma _{1}\rho _{01}-\sigma
_{2}\rho _{02})^{2}\right] (T-t)}} \\
&&+\frac{1}{2}\sqrt{\left[ (\sigma _{1}^{2}+\sigma _{2}^{2}-2\rho
_{12}\sigma _{1}\sigma _{2})-(\sigma _{1}\rho _{01}-\sigma _{2}\rho
_{02})^{2}\right] (T-t)}, \\
d_{2}(x) &=&d_{1}(x)-\frac{1}{2}\sqrt{\left[ (\sigma _{1}^{2}+\sigma
_{2}^{2}-2\rho _{12}\sigma _{1}\sigma _{2})-(\sigma _{1}\rho _{01}-\sigma
_{2}\rho _{02})^{2}\right] (T-t)},
\end{eqnarray*}
$\Phi (\cdot )$ is the standard Gaussian distribution function and $\phi
(\cdot ;\mu ,\nu )$ denote a Gaussian density function with mean $\mu $ and
variance $\nu $.

Clearly, if $\sigma _{1}\rho _{01}=\sigma _{2}\rho _{02}$ (that is, the
covariance between processes $dr(t)$ and $dS_{1}(t)$ is the same as the
covariance between processes $dr(t)$ and $dS_{2}(t)$), then the pricing
formulae for exchange options are the same for both the stochastic and
deterministic term structures. Therefore, it is attempting to argue that we
could use this solution as a control variate if one wants to do Monte-Carlo
simulation to price spread option with nonzero strike price $K$.

\subsection{When underlying assets pay dividends}

The pricing formula for exchange options above when interest rates are
stochastic is derived with the assumption that the two underlying assets,
such as stocks, pay no dividends during the options' life. However, in case
of the underlying assets also pay constant or known dividends, the question
will become the general spread options pricing problem. 

Assume the amount of dividends $d_{1}(t_{1}),$ $\cdots ,$ $d_{1}(t_{m})$ for
the first asset and $d_{2}(s_{1}),$ $\cdots ,$ $d_{2}(s_{k})$ for the second
asset to be paid at the the dates $0<t_{1}<\cdots <t_{m}<T$ and $%
0<s_{1}<\cdots <s_{k}<T$ respectively are known in advance. Without loss of
generality, we can write the dividend streams for both assets by 
\[
d_{i}(t_{1}),\cdots ,d_{i}(t_{n}),\qquad i=1,2
\]
where $n$ is the number of dividends payment dates for both assets.
Therefore, at time $t,$ the present values of all future dividends will be
\[
\sum_{j=1}^{n}d_{i}(t_{j})e^{-\int_{t}^{t_{j}}r(\tau )d\tau
}I_{[t,T]}(t_{j}),\qquad i=1,2,
\]
and the values of all dividends paid after time $t$ and compounded at the
risk-free rate till the option's maturity date $T$ is given by
\[
\sum_{j=1}^{n}d_{i}(t_{j})e^{\int_{t_{j}}^{T}r(\tau )d\tau
}I_{[t,T]}(t_{j}),\qquad i=1,2.
\]
By the same argument of Heath and Jarrow (1988), we decompose the capital
gain processes $G_{i}(t)$ into the asset price process $S_{i}(t)$ and the
dividends streams. Assume further that the $G_{i}(t)$ also follows the
geometric Brownian motions, that is,
\[
\frac{dG_{i}(t)}{G_{i}(t)}=\mu _{i}dt+\sigma _{i}dW_{i}(t),\qquad i=1,\text{ 
}2.
\]
Then the capital gain $G_{i}(t)$ may be written by  
\begin{eqnarray*}
G_{i}(t) &=&S_{i}(t)+\sum_{j=1}^{n}d_{i}(t_{j})e^{\int_{t_{j}}^{T}r(\tau
)d\tau }I_{[t_{j},T]}(t) \\
&=&S_{i}(t)+D_{i}(t)
\end{eqnarray*}
where 
\[
D_{i}(t)=\sum_{j=1}^{n}d_{i}(t_{j})e^{\int_{t_{j}}^{T}r(\tau )d\tau
}I_{[t_{j},T]}(t).
\]
Since the dynamics of the capital gains processes $G_{i}(t),$ $i=1,$ $2,$
under the martingale measure $Q,$ is 
\[
\frac{dG_{i}(t)}{G_{i}(t)}=r(t)dt+\sigma _{i}dW_{i}(t),
\]
and $G_{i}(0)=S_{i}(0)$ and $G_{i}(T)=S_{i}(T)+D_{i}(T),$ $i=1,$ $2.$
Therefore, the risk-neural pricing formula will be
\begin{eqnarray*}
&&E_{t}^{Q}\left[ e^{-\int_{t}^{T}r(s)ds}\ \left[ S_{1}(T)-S_{2}(T)\right]
^{+}\right]  \\
&=&E_{t}^{Q}\left[ e^{-\int_{t}^{T}r(s)ds}\ \left[
G_{1}(T)-G_{2}(T)-(D_{1}(T)-D_{2}(T))\right] ^{+}\right] .
\end{eqnarray*}
In this case, we will have to employ numerical method to value the option
price.

\section{Conclusion}

In this work, we have discussed the issue of pricing exchange options and
spread options. Closed form for the exchange option price is provided
explicitly when the interest rate is stochastic. Our result is valid for
most general term structure model of one factor, which includes Vasicek
model, CIR model, and well-known models as special cases. Our result
indicates that only the correlation coefficients between the interest rate
and the underlying assets will affect the exchange option price. In one
special case, a completely explicit form of the option pricing can be
obtained. We have argued that it is possible to use this solution as a
control variate when doing Monte-Carlo simulation to price spread options
for nonzero strike price $K$ and stochastic interest rate.

\section*{Acknowledgment}

We are indebted to Prof. Phelim P. Boyle, Prof. Don McLeish, 
Prof. James Redekope, Prof. K. S. Tan, Dr. Zejiang 
Yang, Dr. Ti Wang, Dr. Wei Qian, Dr. H. Huang, Dr. W. H. Zou, 
Prof. Kevin Wang, Prof. Ramzi Khuri
for discusions and interactions. Any errors
of this article are solely due to ourselves.

\end{document}